\documentclass[reprint,amsmath,amssymb,aps,]{revtex4-1}

\usepackage{dcolumn}
\usepackage{graphicx}

\begin{document}

\title{Observation of the fine structure for rovibronic spectral 
lines in visible part of emission spectra of $D_2$}

\author{B. P. Lavrov}
\email{lavrov@pobox.spbu.ru}
\author{I. S. Umrikhin}
\author{A. S. Zhukov}
\affiliation{Faculty of Physics, St.-Petersburg State University, \\
Sankt Petersburg, 198504, Russia}

\date{\today}

\begin{abstract}
For the first time in visible part of the $D_2$ emission spectrum the 
pseudo doublets representing partly resolved fine structure of 
rovibronic lines have been observed. They are characterized by 
splitting values about $0.2$ cm$^{-1}$ and relative intensity 
of the doublet components close to $2.0$. It is shown that they 
are determined by triplet splitting in lower rovibronic levels of 
various $^3\Lambda_g^\pm \to c^3\Pi_u^-$ electronic transitions. It is 
proposed to use an existence of such partly resolved fine structure 
patterns for identification of numerous unassigned spectral lines of 
the $D_2$ molecule coming from great variety of triplet "gerade" 
electronic states to vibro-rotational levels of the $c^3\Pi_u^-$ state.
\end{abstract}


\maketitle

The present paper reports new observations concerning triplet-triplet 
electronic-vibro-rotational (rovibronic) spectral lines in visible part 
of the emission spectrum of the $D_2$ molecule. There are several 
peculiarities of current knowledge of triplet rovibronic 
states and radiative transitions between them which make 
it possible to consider this knowledge as insufficient and which motivated 
our experimental studies.

Most of spectral lines in visible and near infrared parts of the 
emission spectrum of molecular deuterium are not classified so far. 
Thus, for example in the latest compilation of experimental data 
\cite{FSC1985} the working list of 27488 recorded lines (within the 
wavelength ranges $\approx 309-1192$ and $1647-2780$ nm) 
contains only 8243 assignments. 
In our opinion it is difficult to consider such situation as normal for an 
isotopomer of simplest neutral molecule (four-particle quantum system).

Almost all experimental data on wavenumbers for triplet rovibronic 
transitions of $D_2$ (3117 lines in \cite{FSC1985}) were obtained 
by means of traditional technique --- photographing an image located 
in a focal plane of long-focus spectrographs \footnote{Non-linear 
response of photographic recording doesn't ensure precise and reliable 
measurements of intensity distributions along the dispersion direction. 
It leads to occurrence of systematic errors in evaluation of wavenumber 
values of blended lines \cite{LU2008, LU2009}. This is a wide-spread 
situation in rather dense multiline spectra of molecular hydrogen 
isotopomers having low mass and large Doppler profiles (e.g. about 
20\% of the $D_2$ triplet lines in \cite{FSC1985}).}. The only 
exceptions are wavenumbers of the 81 rovibronic lines in \cite{DabrHerz} 
and 3 lines in \cite{Davies} obtained in middle IR (about $4.5$ $\mu m$) by 
FTIR (Fourier transform infrared) and laser spectroscopy. 

In contrast to the $H_2$ spectrum where both fine (FS) and 
hyperfine (HFS) structures of triplet lines and levels were 
studied for many electronic states and by various methods (see e.g. 
bibl. in \cite{JOR1994_3}), for the $D_2$ molecule only fragmentary 
data concerning only FS were obtained, namely: fine structure splitting 
values for 11 rovibronic levels of the $d^3\Pi_u^-$ state measured 
by MOMRIE (microwave optical magnetic resonance induced by electrons) method 
\cite{FM1973}; pseudo doublets of partly resolved triplet structure for 18 
from 81 recorded triplet lines in \cite{DabrHerz}, and completely resolved 
FS for 3 rovibronic lines in \cite{Davies}.

Most complete sets of data concerning wavenumber values in visible 
and near IR together with empirical values of rovibronic energy levels 
were reported in \cite{Dieke1972} for the $H_2$ and in 
\cite{FSC1985} for the $D_2$ molecules. They are based on experimental 
results of G.H.~Dieke and co-workers first reported in \cite{Dieke1958}. 
Describing his experimental setup in \cite{Dieke1958} G.H.~Dieke mentioned 
that "In the low pressure, low temperature discharge the lines are 
considerably sharper and for instance the pseudo doublet structure of 
the $2p^3\Pi$ state which is about 0.2 cm$^{-1}$ is well resolved under 
these conditions. This requires a resolving power of 100000 in the visible." 
But in later compilations of the data for both $H_2$ \cite{Dieke1972} and 
$D_2$ \cite{FSC1985} isotopomers the fine structure of lines and empirical 
rovibronic energy levels was not mentioned at all. Moreover, in both cases 
the reported values of experimental errors ("few hundredth cm$^{-1}$" for 
$H_2$ \cite{Dieke1972} and 0.05 cm$^{-1}$ for $D_2$ \cite{FSC1985}) are 
about one order of magnitude smaller than splitting in partly resolved 
fine structure earlier reported in \cite{FR1947} and mentioned in 
\cite{Dieke1958}. It is unclear how empirical rovibronic energy values 
were obtained with such precision when more pronounced effect of the FS 
splitting was not taken into account.

There is noticeable asymmetry in studies of visible and near 
IR spectra of light ($H_2$) and heavy ($D_2$) isotopomers of hydrogen 
molecule. The FS of the $H_2$ spectral lines was discovered by 
O.W.~Richardson and W.E.~Williams as early as in 1931 \cite{RW1931} exactly in 
visible part of the spectrum (see also \cite{FR1947}). Although both 
isotopomers should have similar values of the FS splitting it's 
observation in visible spectrum of the $D_2$ molecule was not reported 
in the literature known to authors.

The goal of present work was to study an opportunity of resolving the fine 
structure in visible spectrum of the $D_2$ molecule by means of spectroscopic 
technique developed in \cite{LU2008, LU2009, LMU2011}. It is based on 
achieving certain level of {\it "optical resolution"} of a spectrograph, 
recording spectral intensity distributions by matrix photoelectric 
detector, and by numerical deconvolution (inverse to the convolution 
operation) of recorded spectra. In dense multiline rovibronic spectra 
of $H_2$ and $D_2$ molecules actual {\it "digital resolution"} achievable 
by our technique could be much higher than {\it "optical resolution"} of 
spectrometer which is limited not by its resolving power but by large 
Doppler broadening of spectral lines (see below). Thus it is possible to 
speak about some kind of sub-Doppler high resolution spectroscopy.

The spectroscopic part of our experimental setup was described in 
\cite{LMU2011}. The $2.65$~m Ebert--type spectrograph with 1800 line/mm 
diffraction grating 100 mm wide was equipped with additional camera lens and 
computer-controlled CMOS matrix detector ($22.2 \times 14.8$ mm$^2$, 
$1728 \times 1152$ triples of the Red, Green and Blue (RGB) 
photo detectors) \footnote{Actually every pixel of the matrix contains
four RGGB detectors, but one from two identical G photodetectors was not
used for recording of spectra.}. The calibrated 
spectrometer makes it possible in the fully automatic regime to record 
sets of individual windows (sections of a spectrum about 1.6 nm wide) 
at the experimentalist's choice, or survey spectra by measuring sequences 
of successive windows with a specified overlap. Thus we obtained digital 
automatic spectrometer with following characteristics: effective focal 
length about 7 meters, linear dispersion $0.077 \div 0.065$ nm/mm (for the 
wavelength region $400 \div 700$ nm). Maximal optical resolving power 
(up to 180000) was achieved in resolving HFS components of Hg lines 
546.1 and 404.6 nm, FWHM of the instrumental profile for 
those wavelengths being 0.021 cm$^{-1}$ and 0.028 cm$^{-1}$ respectively.

When the entrance slit is uniformly illuminated, signals of one type photo 
detectors (B, G, or R), located in the same vertical column, 
carry information concerning the brightness of the radiation at the same 
wavelength. Averaging makes it possible to increase the sensitivity of 
the spectrometer and the signal-to-noise ratio (SNR) for data obtained by 
single exposure. To reach required value of SNR we made many records (up 
to 150 shots for the same wavelength fragment and the same discharge 
conditions). Averaging of those results made it possible to reach SNR 
value up to $10^4$ (see \cite{LMU2011}).

For recording the $D_2$ spectra with low resolution,
high sensitivity and large population
of high rotational levels we used hot-cathode capillary-arc discharge
lamp LD-2D described in \cite{GLT1982} (pure $D_2$ under pressure 
$\approx 6$ Torr, capillary inner diameter \O{2} mm, current 
density $\approx 10$ A/cm$^2$). Gas temperature $T = 1890 \pm 170 $ K
was obtained from the intensity distribution in the rotational structure of
the $(2-2)$ Q-branch of Fulcher-$\alpha$ band system (see e.g. 
\cite{L1980, AKKK1996}). It corresponds to Doppler linewidths (FWHM) 
$\Delta \nu_D$ = $0.22 \div 0.37$ cm$^{-1}$ for $1/\nu = 420 \div 700$ nm.
Therefore we were able to open the entrance slit of the spectrometer up to 
60 $\mu$m for gaining more signal (and corresponding decrease in data 
accumulation time) without significant loss in resolution.

To achieve best possible optical resolution we have to decrease 
Doppler broadening by diminution of the gas temperature in plasma. 
It is obviously favorable for increasing spectral
resolution, but lowering the temperature automatically leads to lower
population densities of high rotational levels in ground and exited
electronic states and to much smaller intensities of corresponding
spectral lines. Therefore we had to use some compromise
plasma conditions. Thus in high resolution experiments we used 
glow discharge with cold cathode and water cooled walls.
Additional third electrode with the axial cylindrical hole \O{4} mm
was located on discharge axes between cathode and anode.
Through a hole in an anode the flux of radiation emitted by plasma inside
additional electrode was focused on the entrance slit of the spectrometer.
With this geometry we got current density as low as 0.4 A/cm$^2$ and
$T = 610 \pm 20$ K.

It should be emphasized once more that the overwhelming majority of data 
on the wavenumbers for rovibronic transitions of the $D_2$ molecule is 
obtained by photographic recording of spectra up to now 
(see e.g. \cite{RLTBjcp2006, RLTBjcp2007}). Our way of determining 
wavenumber values is based on linear response of CMOS matrix photo detector
on the spectral irradiance and digital intensity recording. Both things 
provide an extremely important advantage of our technique over traditional 
photographic recording with microphotometric or visual comparator reading. 
It not only makes it easier to measure the relative 
spectral line intensities but also makes it possible to investigate the 
shape of the individual line profiles and, in the case of overlap of the 
contours of adjacent lines (so-called blending), to carry out numerically 
the deconvolution operation and thus 
to measure the intensity and wavelength of blended lines. As 
is well known, it is this blending that makes it very hard to analyze 
dense multiline spectra of the $D_2$ molecule \cite{FSC1985}.

\begin{figure*}
\includegraphics{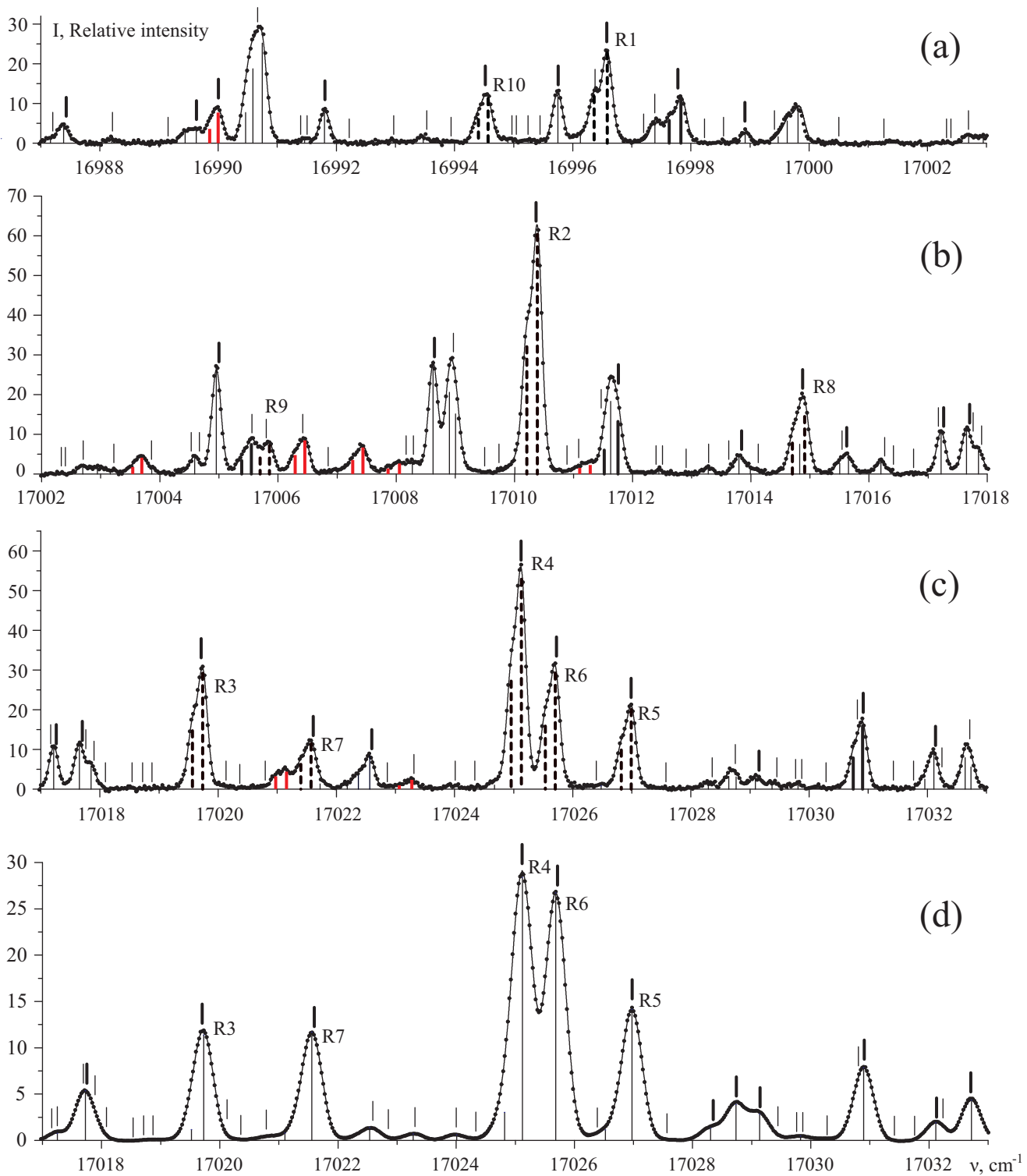}
\caption{\label{fig:spectra}Part of the $D_2$ spectrum containing first 10 lines 
of the R-branch for $(1-1)$ band of the 
$i^3\Pi_g^- \to c^3\Pi_u^-$ electronic transition obtained in high (a, b, c)
and low (d) resolution experiments. Experimental data $I_k^{expt}$ in relative 
units are shown by circles. Solid line represents the intensity distribution 
calculated as a sum of optimal Voigt profiles. Wavenumber values reported in 
\cite{FSC1985} are marked by short vertical line segments above $I_k^{expt}$,
those of them shown in bold were used for calibration of our spectrometer.
Spectral lines obtained by the deconvolution are presented as "stick diagrams" 
indicating their wavenumber positions $\nu_i^0$ and amplitudes $A_i$. Pairs of 
sticks painted in bold represent the pseudo doublets having specific 
characteristics (see text).}
\end{figure*}

We are treating the problem of wavenumber determination as that of conditional
optimization: parameterization of a model for an intensity distribution 
and determination of optimal set of parameters by searching a global minimum
of an objective function under specified conditions.
Thus for small regions of the spectrum ($\approx 0.5$ nm wide, about 
one-third of a window, containing $500 \div 600$ vertical columns 
of photodetectors, see Fig.~\ref{fig:spectra}) the observed spectral 
intensity distribution (dependence of the photoelectric signal of k-th 
photodetector $I_k^{expt}$ on the detector position $x_k$) was approximated 
by a superposition of a finite number $M$ of line profiles $f_i(x)$ 
with a width $\Delta x$ common for all lines within an analyzed region:
\begin{equation}
I^{calc}(x) = I_{bg} + 
  \sum\limits_{i=1}^{M} A_i f_i(x, x^0_i, \Delta x), \label{eq:sumprofile}
\end{equation}
where $I_{bg}$ is a constant background intensity, $A_i$ -- amplitude 
of i-th profile (intensity in the line center $x^0_i$).

We used an objective function in the form of a sum of squares of deviations 
between experimental and "synthesized" intensity distributions 
\begin{equation}
\Phi(\{A_i, x^0_i\}_{i=1...M}, \Delta x, I_{bg}) = 
  \sum\limits_{k=1}^{K} (I_k^{expt} - I^{calc}(x_k))^2, \label{eq:r2}
\end{equation}
where $K$ indicates a number of experimental intensity values $I_k^{expt}$ 
in the spectral region under the study. 

\begin{table*}
\caption{\label{tab:nu}
Wavenumber values (in cm$^{-1}$) and relative intensities $I_S/I_W$ 
for strong and weak components for pseudo doublets of the R-branch 
lines for the $(1-1)$ band of the $i^3\Pi_g^- \to c^3\Pi_u^-$ 
electronic transition (the T- 3e-2c (1-1) RN" rovibronic transitions in G.H.Dieke 
notation \cite{FSC1985}). The $\nu_S$ and $\nu_W$ are wavenumbers of 
strong and weak components; $\Delta \nu_{SW} = \nu_S - \nu_W$; 
$\Delta \nu_{LR}$ -- the wavenumbers obtained in low resolution experiments.
Experimental errors (one SD) are shown in brackets in unites of last significant digit.}

\begin{ruledtabular}
\begin{tabular}{ld@{}dddd}
\multicolumn{2}{c}{\cite{FSC1985}} & \multicolumn{4}{c}{Present work} \\ \cline{1-2} \cline{3-6}
Assignment & \multicolumn{1}{c}{$\nu$} & \multicolumn{1}{c}{$\nu_{LR}$} & 
\multicolumn{1}{c}{$\nu_S$, $\nu_W$} & \multicolumn{1}{c}{$\Delta \nu_{SW}$} & 
\multicolumn{1}{c}{$I_S/I_W$} \\ \hline
T- 3e-2c (1-1) R1  & 16996.58 & 16996.57(2)  &  16996.58(3)s & 0.22(5) & 1.93(5)  \\ 
                   & 16996.38 & 16996.31(2)  &  16996.36(4)w &         &          \\
T- 3e-2c (1-1) R2  & 17010.36 & 17010.37(2)  &  17010.39(3)s & 0.18(4) & 1.88(2)  \\ 
                   &          &              &  17010.21(3)w &         &          \\
T- 3e-2c (1-1) R3  & 17019.71 & 17019.73(2)  &  17019.74(3)s & 0.17(5) & 1.99(4)  \\ 
                   &          &              &  17019.57(4)w &         &          \\
T- 3e-2c (1-1) R4  & 17025.11 & 17025.12(2)  &  17025.13(3)s & 0.17(4) & 1.94(3)  \\ 
                   &          &              &  17024.96(3)w &         &          \\
T- 3e-2c (1-1) R5  & 17026.97 & 17026.98(2)  &  17026.99(3)s & 0.17(5) & 2.05(7)  \\ 
                   &          &              &  17026.82(4)w &         &          \\
T- 3e-2c (1-1) R6  & 17025.71 & 17025.69(2)  &  17025.70(3)s & 0.17(5) & 1.84(4)  \\ 
                   &          &              &  17025.53(4)w &         &          \\
T- 3e-2c (1-1) R7  & 17021.60 & 17021.56(2)  &  17021.57(4)s & 0.18(6) & 1.82(11) \\ 
                   &          &              &  17021.40(4)w &         &          \\
T- 3e-2c (1-1) R8  & 17014.87 & 17014.88(2)  &  17014.91(4)s & 0.21(6) & 1.90(60) \\
                   &          &              &  17014.70(4)w &         &          \\
T- 3e-2c (1-1) R9  & 17005.79 & 17005.82(2)  &  17005.86(4)s & 0.16(6) & 1.80(20) \\ 
                   &          &              &  17005.70(5)w &         &          \\
T- 3e-2c (1-1) R10 & 16994.53 & 16994.54(2)  &  16994.57(4)s & 0.16(6) & 1.89(14) \\ 
                   &          &              &  16994.40(4)w &         &          \\
\end{tabular}
\end{ruledtabular}
\end{table*}

If the experimental errors 
of the $I_k^{expt}$ values are random and distributed according to a 
normal (Gaussian) law, the solution obtained by the least-squares 
criterion for (\ref{eq:r2}) corresponds to the maximum likelihood principle.
For determining a global minimum of the objective function 
(\ref{eq:r2}) in multidimensional parameter 
space we used special computer program based on Levenberg-Marquardt's 
algorithm \cite{L1944, M1963}. 

Our studies showed that in HFS of Hg spectral lines 
\cite{LMU2011} and in the low resolution experiments
approximation of line profiles by Gaussian function 
$f_i(x) = exp \left( -\frac{1}{2} [x - x^0_i]^2 / \Delta x_G^2 \right)$
(with the linewidth $\Delta x_G$) was adequate providing random scatter 
of the $I_k^{expt} - I^{calc}(x_k)$ deviations and high enough accuracy. 

The analysis of line profiles obtained in high resolution experiments 
showed that Gaussian function is insufficient and we had to use more 
flexible Voigt profiles:
\begin{equation}
f_i(x) = \frac
  { \displaystyle 
    \int\limits_{-\infty}^{+\infty}
    \frac{exp(-t^2)}{ \left( \frac{\Delta x_L}{\sqrt{2}\Delta x_G} \right)^2 +
    \left( \frac{x-x^0_i}{\sqrt{2}\Delta x_G} - t\right)^2} dt }
  { \displaystyle
    \int\limits_{-\infty}^{+\infty}
    \frac{exp(-t^2)}{ \left( \frac{\Delta x_L}{\sqrt{2}\Delta x_G} \right)^2 + 
  t^2} dt }, \label{eq:voigt}
\end{equation}
where $\Delta x_L$ --- Lorentzian linewidth. 

The values for all the $2M + 3$ parameters
($\{A_i, x^0_i\}_{i=1...M}$, $\Delta x_G$, $\Delta x_L$, and $I_{bg}$)
obtained by minimizing (\ref{eq:r2}) are optimal for the observed intensity 
distribution under the condition of identical $\Delta x_G$ and 
$\Delta x_L$ values for all the lines. Thus it is possible to obtain
optimal values of the amplitude and a line center for each spectral line 
as well as common value of total "observed" line width $\Delta x$ calculated 
from optimal values of $\Delta x_G$ and $\Delta x_L$.

In the case of long-focus spectrometers the dependence of the wavelength 
on the coordinate along direction of dispersion is close to linear in the 
vicinity of the center of the focal plane. It can be represented as a power 
series expansion over of the small parameter $x/F$ (The $x$-coordinate
represents small displacement from the center of the matrix detector, $F$ is 
the focal length of the spectrometer), which in our case does not exceed 
$2 \times 10^{-3}$ \cite{LMU2011}. On the other hand, the wavelength dependence 
of the refractive index of air $n(\lambda)$ is also close to linear inside 
a small enough part of the spectrum. Thus, when recording narrow spectral 
intervals, the product $\lambda_{vac}(x) = \lambda(x) n(\lambda(x))$ has the 
form of a power series of low degree. This circumstance makes it possible to 
calibrate the spectrometer directly in vacuum wavelengths 
$\lambda_{vac} = 1/\nu$, thereby avoiding the technically troublesome problem 
of accurate measuring the refractive index of air for various experimental 
conditions.
For spectrometer calibration the experimental vacuum wavelength values ($1/\nu$) 
of bright, non-blended $D_2$ lines from \cite{FSC1985} were used as the standard 
reference data \footnote{Our previous studies \cite{LMU2011} of the emission 
spectrum of capillary-arc discharge lamp analogous to that described in 
\cite{LSh1979} but filled with the $D_2 + H_2 + Ne$ gas mixture show that 
wavenumber values of atomic lines of $Ne$ 
\cite{SS2004Ne} and those of bright non-blended lines of the $H_2$ 
\cite{Dieke1972} and $D_2$ \cite{FSC1985} molecules are in rather good mutual 
accordance and may be used as reference data for spectrometer calibration 
providing an accuracy about 10$^{-3}$ nm \cite{LU2011}.}.
They show small random spread around 
smooth curve representing dependence of the wavelengths on positions of 
corresponding lines in the focal plane of the spectrometer. Moreover these 
random deviations are in good accordance with normal distribution. 
Thus it is possible to obtain precision for new wavenumber values 
better than that of the reference data due to smoothing. The calibration curve 
of the spectrometer was obtained by polynomial least-squares fitting the 
data with accuracy better than $2 \times 10^{-3}$ nm. 

Following the way described above we measured the $D_2$ spectra in low and high 
resolution experiments for wavenumber regions $14378.80 \div 23894.65$ cm$^{-1}$ 
($695 \div 418$ nm) and $15948.82 \div 18331.28$ cm$^{-1}$ ($627 \div 545$ nm) 
\cite{LU2011} respectively. Within these intervals 11986 and 3518 spectral lines 
were distinguished after the deconvolution. Detailed analysis of the data will 
be reported in subsequent papers. In the present short communication we shall 
restrict ourselves to consideration of one particular, but rather typical case 
which allow to illustrate some general features of our first observations.

As an example four fragments of the $D_2$ spectrum containing first 10 lines of 
the R-branch for $(1-1)$ band of the $i^3\Pi_g^- \to c^3\Pi_u^-$ electronic 
transition are shown in Fig.~\ref{fig:spectra}. First fragments ($a, b, c$) are 
three parts of the same window recorded in the high resolution experiment 
(discharge current I = 30 mA, entrance slit $\Delta S$ = 15 $\mu$m, observed 
FWHM $\Delta \nu$ = 0.18 cm$^{-1}$). They were used separately in the 
deconvolution procedure described above. Fourth fragment ($d$) is identical to 
the third one ($c$), but it was obtained in low resolution experiment (I = 300 
mA, $\Delta S$ = 60 $\mu$m, $\Delta \nu$ = 0.39 cm$^{-1}$). One may see that two 
identical fragments measured with different spectral resolution are 
qualitatively different. In the low resolution case all lines look like singles 
having symmetrical profiles. In high resolution experiments the partly resolved 
fine structure of some lines becomes apparent as asymmetry of their profiles,
although some other lines remain single with symmetric profiles 
(see Fig.~\ref{fig:spectra}($a$, $b$)). 
This is a result of say {\it "optical resolution"} only. The deconvolution of 
measured intensity distributions based on numerical optimization technique 
described above provides an opportunity to recognize narrow substructures 
within observed asymmetric profiles. The results of such 
{\it "digital resolution"} are shown in Fig.~\ref{fig:spectra} as 
"stick diagrams" of the individual components indicating their wavenumber 
positions $\nu_i^0$ and amplitudes $A_i$. Numerical 
data concerning the $(1-1)$ $R-$branch lines under the study are presented 
also in Tab.~\ref{tab:nu}. One may see from the table that in our conditions 
this technique is able to provide high enough precision in wavenumbers and 
relative intensities of latent spectral lines. Moreover, additional resolving 
power obtained by the deconvolution is sufficiently higher than that 
corresponding to Rayleigh criterion.

Among many lines shown in Fig.~\ref{fig:spectra}($a$, $b$, $c$) the 22
pairs of recognized lines (sticks) catch one's eye, because they have
distinguishing features: the splitting value is about $0.2$ cm$^{-1}$,
and intensity ratios of the violet (strong) and red (weak) components 
are close to $2.0$. The sticks representing such pseudo doublets are
shown in bold. 10 of them (painted as black dash sticks) were previously
classified as single rovibronic lines belonging to the $(1-1)$ $R-$branch
of the $i^3\Pi_g^- \to c^3\Pi_u^-$ electronic transition \cite{FSC1985}.
4 other cases (painted black) were also classified as triplet lines
coming to vibro-rotational levels of the $c^3\Pi_u^-$ state \cite{FSC1985}.
Recently these assignments were confirmed by statistical analysis of the
experimental wavenumbers in the framework of Rydberg-Ritz combination
principle \cite{LU2008}. 8 other pairs (painted in red) are not assigned
so far.

\begin{figure}
\includegraphics{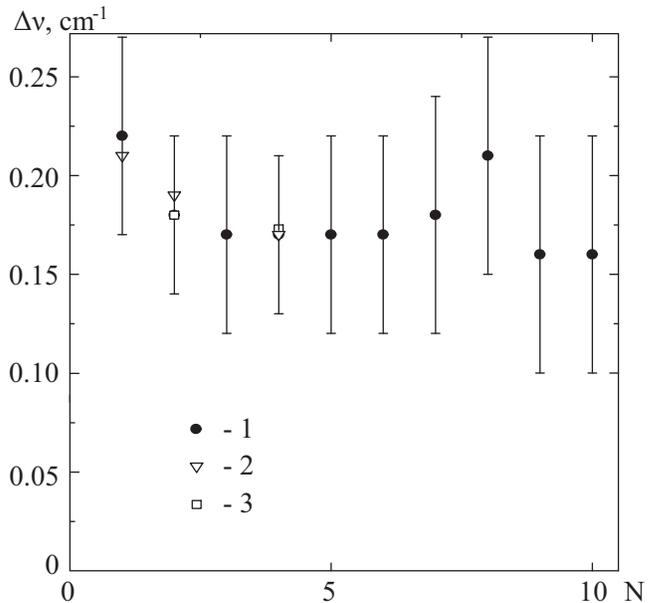}
\caption{\label{fig:splitting}
Values of the splitting in pseudo doublets observed in the present work for the
$i^3\Pi_g^-, v' = 1, N'' + 1 \to c^3\Pi_u^-, v'' = 1, N''$ rovibronic 
transitions of $D_2$ (points 1) and those for the 
$a^3\Sigma_g^+, v' = 2, N'' \to c^3\Pi_u^-, v'' = 1, N''$ transitions of $D_2$ 
obtained by FTIR spectroscopy in \cite{DabrHerz} (points 2).
3 -- the splitting values in pseudo doublets of the 
$c^3\Pi_u^-, v = 1, N = 2, 4$ rovibronic levels of the $H_2$ molecule, 
calculated from the data reported in \cite{JOR1994_2}.}
\end{figure}

The splitting values obtained in present work for the $(1-1)$ $R-$branch
lines are shown in Fig.~\ref{fig:splitting} together with analogous data
obtained in middle IR for the $a^3\Sigma_g^+ \to c^3\Pi_u^-$ transitions
of $D_2$. One may see that the results obtained from two different band
systems having common $c^3\Pi_u^-, v'' = 1, N''$ rovibronic states are
in good agreement. Moreover, observed in both experiments with $D_2$
molecule the wavenumber splittings are almost the same as pseudo doublet
splitting of the FS sublevels of the $c^3\Pi_u^-, v'' = 1, N'' = 2, 4$
levels of the $H_2$ molecule.

Thus it is natural to interpret all observed pseudo doublets as partly
resolved FS patterns of the 
$^3\Lambda_g^\pm, v', N' \to c^3\Pi_u^-, v'', N''$ rovibronic transitions
mainly determined by FS splitting of rovibronic levels in the $c^3\Pi_u^-$
state (Here $\Lambda$ is quantum number for projection of electronic orbital 
angular momentum onto internuclear axes, $v$ --- vibrational quantum number,
$N$ --- quantum number of total angular momentum excluding electron and nuclear
spins, and upper and lower states being marked by single and double primes
respectively).

The intensity ratios $I_s / I_w$ for the $(1-1)$ $R-$branch lines under
the study are listed in Tab.~\ref{tab:nu}. These values can't be compared
with any other experimental results because they were not reported in
the literature known to us. At the same time one may see that our experimental
values are close to $2.0$. Exactly this value may be obtained by well-known 
Burger-Dorgello-Ornstein sum rule for intensities within narrow multiplets 
when one assumes that the triplet splitting in upper rovibronic states may 
be neglected while in the lower rovibronic states $c^3\Pi_u^-, v = 1, N''$ two 
fine structure sublevels ($J'' = N'' - 1$ and $J'' = N'' + 1$) are close to 
each other and located noticeably lower than that with $J'' = N''$ 
\footnote{There is 6 possible sequence orders of FS sublevels leading 
to formation of visible pseudo doublet structure of lines and levels.
Only one of them formulated above provides the intensity ratio 
$I_s / I_w = 2.0$. The 4 other orders give strong dependence of $I_s / I_w$
value on $N''$, and another one gives $I_s / I_w = 1/2$.}.
These assumptions are in agreement with IR tunable laser observations 
($E_{J'' = 2} \approx E_{J'' = 0} < E_{J'' = 1}$) for the fine structure of 
the  $a^3\Sigma_g^+, v' = 4, N' = 3 \gets c^3\Pi_u^-, v'' = 3, N'' = 1$ 
rovibronic transition reported in \cite{Davies}. Thus our ability to measure 
both the intensities and splitting values gives us an opportunity to get 
information about an order and separation of the fine structure sublevels. 

Finally, two main results of our observations may be formulated as follows.
The deconvolution of intensity distributions recorded by a matrix photoelectric 
detector by means of numerical optimization procedure is a powerful tool for 
determining wavenumbers and intensities of substructures within apparent line 
profiles masked by overlapping of adjacent lines (blending) and line broadening 
in traditional photographic recording of spectra. In contrast to fragmentary 
results of tunable laser techniques, such Doppler-free classic spectroscopy 
is able to provide huge volumes of data for broad regions of molecular 
spectra. It should be stressed that we are working in visible part of the 
spectrum, most suitable for various applications.
Even partly resolved 
fine structure of spectral lines provides an opportunity to expand the existing 
identification of triplet rovibronic lines by detecting those doublets in 
experimental spectra. The doublets analyzed above are especially promising 
because they are easily recognizable in the spectrum due to their distinguishing 
features. Within the spectral region under the study ($545 \div 627$ nm) we 
already found more than $200$ pairs of unassigned lines which may represent 
pseudo doublets of partly resolved FS of rovibronic transitions between 
$^3\Lambda_g^\pm$ and $c^3\Pi_u^-$ electronic states of the $D_2$ molecule \cite{LU2011}. 

\begin{acknowledgments}
The authors are indebted to S.C.~Ross for providing the electronic version of 
the appendix C from \cite{FSC1985}. Present work was supported, in part, by 
Russian Foundation for Basic Research, Grant No. 10-03-00571-a.
\end{acknowledgments}

\bibliography{2012LUZ_PhRevA_f}

\end{document}